\documentstyle[11pt,moriond,epsfig]{article}

\def\be{\begin{equation}}
\def\ee{\end{equation}}
\def\bea{\begin{eqnarray}}
\def\eea{\end{eqnarray}}

\newcommand{\gvec}[1]{\hbox{\boldmath$#1$\unboldmath}}
\newcommand{\nvec}[1]{|\gvec{#1}|}
\newcommand{\ie}{\textit{i.e.} }

\begin{document}

\vspace*{4cm}

\vspace*{4cm}

\title{NEUTRINO-OXYGEN INTERACTIONS: ROLE OF NUCLEAR PHYSICS IN THE
ATMOSPHERIC NEUTRINO ANOMALY}

\author{ J. MARTEAU, J. DELORME, M. ERICSON\,\footnote{Also at Theory Division, CERN, CH-1211 Geneva 23, Switzerland} }

\address{Institut de Physique Nucl\'eaire de Lyon,\\ 
IN2P3-CNRS et Universit\'e Claude Bernard-Lyon I,\\ 43, bd du 11 novembre 1918, 69622
Villeurbanne, France}

\maketitle\abstracts{
The apparent anomaly in the ratio of muon to electron atmospheric neutrinos  
first observed by Kamiokande and IMB has been confirmed by Super-Kamiokande and Soudan-2. The experimental analysis, including the asymmetry in  the zenithal distributions of the $ \mu-\mathrm{type} $ events in Super-Kamiokande gives a strong support to the neutrino oscillation hypothesis to solve the anomaly.\\
In this work we are interested by the role of nuclear physics in the neutrino-oxygen reactions used to detect the atmospheric neutrinos. We point out that multi-nucleon excitations of {\it np-nh} type and that nuclear correlations could modify an experimental analysis {\it \`a la} Super-Kamiokande because they lead to a substantial enhancement of the number of {\it 1 \v{C}erenkov ring} retained events.}  

\newpage

\section{Introduction}     
     
The apparent anomaly in the numbers of $\nu_\mu$ and $\nu_e$ produced by interactions of cosmic rays in the atmosphere has been confirmed by several experiments \cite{kamiokande,superkamiokande,imb,soudan}. While the expected ratio of $\nu_\mu$ to $\nu_e$ should be of the order of two in favour of the muonic neutrinos, as indicated by simple arguments on the decays of pions and kaons and confirmed by several theoretical calculations \cite{barr/gaisser/stanev,gaisser/honda/etal}, the measured ratio is significantly lower. Indeed the double ratio:
\begin{equation}
r = \frac{(R_{\mu/e})_{exp}}{(R_{\mu/e})_{MC}}
\end{equation}
between the experimental and the simulated $\nu_\mu$ to $\nu_e$ ratios $(R_{\mu / e})$ is found to be of the order of $\sim$ 0.6.  
Furthermore the Super-Kamiokande collaboration has observed an asymmetry in the zenithal distributions of the $\nu_\mu$ events (both sub- and multi- GeV) \cite{superkamiokande}. It appears that the number of upward-going $\nu_\mu$ (\ie coming from the antipodes) is smaller than expected while the number of downward-going $\nu_\mu$ corresponds to the expectations. Together with the symmetric behaviour of the zenithal distributions of the $\nu_e$ events and the results of the Chooz experiment which excludes the possible $\nu_\mu \longrightarrow \nu_e$ solution to the Super-Kamiokande results, this favours a solution of the atmospheric neutrino problem in the $\nu_\mu \longrightarrow \nu_X$ oscillations channels, with $\nu_X = \nu_\tau$ or $\nu_X = \nu_s$ (some kind of sterile neutrino).\\

\section{Role of nuclear physics}     
     
In this work we are interested in the experiments using large underground water \v{C}erenkov detectors: Kamiokande and Super-Kamiokande, IMB. The analysis is based on events where only one \v{C}erenkov ring is detected. These "one \v{C}erenkov ring" (1 \v{C}.R.)
events are usually assumed to be produced by quasi-elastic charged current interactions in which a charged lepton is emitted above \v{C}erenkov threshold and leads to the \v{C}erenkov ring. The nucleon which is ejected from the nucleus is in general below threshold and therefore does not produce another ring (in water the threshold kinetic energy for a nucleon is about 0.5 GeV and the assumption is rather correct). However we know that the nuclear dynamics is far more complex than this simple picture and there are others sources of 1 \v{C}.R. events \cite{marteau}.\\
Indeed the region of energy transfer in processes involving atmospheric neutrinos of $ \sim $ 1 GeV extends from the quasi-elastic peak to the Delta resonance region. The quasi-elastic region provides the most important part of the 1 \v{C}.R. events with the previous assumption on the emitted nucleon. At higher energy we know from electron scattering that {\it 2 particles - 2 holes} excitations are necessary to fill the intermediate region (the so-called {\it dip} region) between the quasi-elastic and the Delta resonance peaks. In the context of atmospheric neutrino such excitations lead to the emission of two nucleons and one charged lepton. Assuming that the nucleons are under threshold one still gets an 1 \v{C}.R. event. The same arguments are valid for  
{\it n particles - n holes} excitations.\\ 
At still higher energy we evaluate the nuclear responses in the Delta resonance region.
If the Delta decays into a pion and a nucleon and if the pion escapes from the nuclear medium above threshold ($ \sim $ 70 MeV in water for a pion), then one gets two \v{C}erenkov rings (one for the charged lepton and the second for the emitted pion). Such events are rejected by the experimental cuts. However the pion in the nucleus is a quasi-particle with a broad width and can decay for instance into a particle-hole excitation. Therefore the decay of a Delta in the nuclear medium can lead to a nucleon and a \textit{nucleon-hole} state. Such a process, where two nucleons are ejected from the nucleus, just produces one \v{C}erenkov ring.\\
The previous arguments lead to the conclusion that a full calculation of the neutrino-oxygen cross sections beyond the quasi-elastic assumption with identification of the possible final states is necessary. This procedure gives the total 1 \v{C}.R. events yields in the atmospheric neutrinos experiments which have to be compared to the quasi-elastic 1 \v{C}.R. events yields usually retained in the simulations. Note that our investigation relies on very simplifying assumptions concerning the particles in the final state. In particular we will neglect throughout this paper the role of pion re-interactions and absorption, the emission of nucleons above \v{C}erenkov threshold and the possible particle misidentification problems. This last assumption is due to a lack of information on the experimental detection efficencies.

\section{Nuclear responses and differential cross sections}    
     
The starting point of this calculation is the inclusive charged-current cross section for the reaction: $ \nu_l \, (\bar{\nu}_l) + ^{16}\hbox{O} \longrightarrow l^- \, (l^+) + X $,
\begin{eqnarray} \label{eq:1}
\frac{\partial^2\sigma}{\partial\Omega \partial k^\prime} & = & \frac{G_F^2 \, \cos^2\theta_c \, \gvec{k^\prime}^2}{2 \, \pi^2} \, \cos^2\frac{\theta}{2} \, \left[ G_E^2 \, (\frac{q_\mu^2}{\gvec{q}^2})^2 \, R_\tau^{NN} \right. \nonumber \\ & + & G_A^2 \, \frac{( M_\Delta - M )^2}{2 \, \gvec{q}^2} \, R_{\sigma\tau (L)}^{N\Delta} \, + \, G_A^2 \, \frac{( M_\Delta - M )^2}{\gvec{q}^2} \, R_{\sigma\tau (L)}^{\Delta\Delta} \nonumber \\ & + & ( G_M^2 \, \frac{\omega^2}{\gvec{q}^2} + G_A^2 ) \, ( - \frac{q_\mu^2}{\gvec{q}^2} + 2 \tan^2\frac{\theta}{2} ) \, ( R_{\sigma\tau (T)}^{NN} + 2 R_{\sigma\tau (T)}^{N\Delta} + R_{\sigma\tau (T)}^{\Delta\Delta} ) \nonumber \\ & \pm & \left. 2 \, G_A \, G_M \, \frac{k + k^\prime}{M} \, \tan^2\frac{\theta}{2} \, ( R_{\sigma\tau (T)}^{NN} + 2 R_{\sigma\tau (T)}^{N\Delta} + R_{\sigma\tau (T)}^{\Delta\Delta} ) \right] 
\end{eqnarray}
where $ G_F $ is the weak coupling constant, $ \theta_c $ the Cabbibo angle, $ k $ and $ k^\prime $ the initial and final lepton momenta, $ q_\mu = k_\mu - k_\mu^\prime = ( \omega,\gvec{q} ) $ the four momentum transferred to the nucleus, $ \theta $ the scattering angle, $ M_\Delta $ ($ M $) the Delta (nucleon) mass. The plus (minus) sign in eq. (\ref{eq:1}) stands for the neutrino (anti-neutrino) case. In a provisional approximation we have neglected in eq. (\ref{eq:1}) the lepton masses and we have kept the leading terms in the development of the hadronic current in $ p/M $, where $ p $ denotes the initial nucleon momentum. The electric, magnetic and axial form factors are taken in the standard dipole parameterization. We have introduced the inclusive \textit{isospin} ($ R_\tau $), \textit{spin-isospin longitudinal} ($ R_{\sigma\tau (L)} $) and \textit{spin-isospin transverse} ($ R_{\sigma\tau (T)} $) nuclear responses functions corresponding to the isospin operator ($\tau_\alpha$) and to the different projections of the spin-isospin operators ($\gvec{\sigma} \,\, \tau_\alpha$) on the direction of the momentum transfer ($\gvec{q}$). Assuming that $O_\alpha^P(j)$ is one of the previous operator, responsible for the coupling of the incident neutrino to a \textit{Particle-hole} state, the corresponding nuclear response is given by: 
\begin{equation} \label{eq:2}
R_\alpha^{PP^\prime} = \sum_n \, \langle n | \sum_{j=1}^A \, O_\alpha^P(j) \, e^{ i \, \gvec{q}.\gvec{x}_j } | 0 \rangle \, \langle n | \sum_{k=1}^A \, O_\alpha^{P^\prime}(k) \, e^{ i \, \gvec{q}.\gvec{x}_k } | 0 \rangle^* \delta(\omega - E_n + E_0)  
\end{equation}
The challenging task is the evaluation of the various nuclear responses. The main problem arising is the knowledge of the nuclear excited states for transfer energies of several hundred MeV. We avoid this difficulty with the model developped by Delorme and Guichon \cite{delorme/guichon} which is based on a semi-classical approximation. This implies the use of local Fermi momentum $ k_F(r) $ which is calculated by the means of an experimental nuclear density: $ k_F(r) = ( 3/2 \, \pi^2 \, \rho(r) )^{1/3} $ and this leads to the "bare" nuclear responses $ R^0(\omega,\nvec{q}) $ (in the following "bare" will mean that the nuclear correlations are switched off). Indeed until now we have considered that the {\it particle-hole} excitations in the medium do not interact with each other. This is of course not true. To properly take this effect into account we adopt a phenomenological parameterization of the interaction between excitations in each reaction channel: a pure contact force with a Landau-Migdal parameter $f^\prime$ is taken in the {\it isospin} channel and a $(\pi + \rho)$-exchange plus a contact force with a Landau-Migdal parameter $g^\prime$ in the {\it spin-isospin} channel. Once the form of this potential is known, we exactly solve the RPA equations in the {\it ring approximation}. This gives the full RPA nuclear responses $ R(\omega,\nvec{q}) $.\\
The modified Delta width in the nuclear medium is split into the contributions of different decay channels \cite{oset/salcedo}: the "quasi-elastic" channel, $ \Delta \longrightarrow \pi \, N $, modified by the Pauli blocking of the nucleon and the distorsion of the pion, the two-body (\textit{2p-2h}) and three-body (\textit{3p-3h}) absorption channels. The (\textit{2p-2h}) excitations which are not already included in the modified Delta width are evaluated by extrapolating the calculations of two-body pion absorption at threshold \cite{shimizu/faessler}. These parameterizations lead to a good description of the pion-nucleus reactions.\\ 
The inclusive nuclear response functions are split into their different contributions. By construction, the "bare" response is the sum of the quasi-elastic, the {\it n particles - n holes} ({\it n = 2,3}) and the resonant Delta ($\Delta \longrightarrow \pi \, N$) responses. In the RPA case we will recognize the same partial contributions supplemented by the {\it coherent pion emission} channel which arises when an intermediate pion exchanged between excited states is placed on its mass-shell. The full response can then be written as:
\begin{equation}
R(\omega,q) = R_{qel}(\omega,q) + R_{np-nh}(\omega,q) + R_{res}(\omega,q) + R_{coh}(\omega,q)   
\end{equation}
We recall that within our model $ R_{qel} $ and $ R_{np-nh} $ will induce 1 \v{C}.R. events and that $ R_{res} $ and $ R_{coh} $ will lead to the emission of 2 \v{C}erenkov rings.\\

\begin{figure}[ht]
\begin{center}
\epsfig{figure=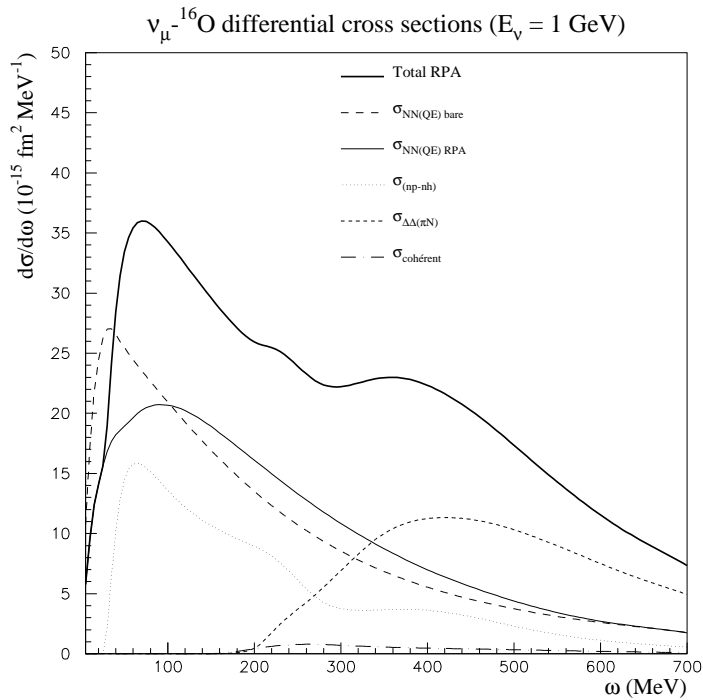,width=10cm,height=10cm}
\end{center}
\caption{Differential charged-current $ \nu_e- ^{16}\hbox{O} $ interactions cross-section versus the energy transfer. The thick curve represents the inclusive cross-section. The following exclusive contributions to the inclusive cross-section are displayed: quasi-elastic (full thin curve), {\it (np-nh)} (dotted curve), resonant Delta ($\Delta \longrightarrow \pi N $) (short dashed curve). Also shown are the "bare" (without RPA correlations) quasi-elastic (long dashed curve) and the coherent pion emission cross-sections (dot-dashed curve).
\label{fig:1}}
\end{figure}

The results for the differential cross section \mbox{$ \partial\sigma / \partial k^\prime $}, which is obtained from the doubly differential cross section (eq. (\ref{eq:1})) by a numerical integration over the solid angle, are shown on figure (\ref{fig:1}). The neutrino energy is fixed at the typical value $E_\nu = 1$ GeV.

The inclusive cross section is given by the thick curve. It gets its main contribution from the quasi-elastic channel (thin full line) which peaks at relatively low energy transfer. For a sake of comparison we have shown the contribution of the quasi-elastic channel without RPA (thin long dashed line). We observe that the cross section is reduced and hardened in the RPA case. The shift in strength is due to the repulsive RPA-interaction in the spin-isospin transverse response. Indeed the Landau-Migdal interaction is repulsive for all values of the transfer and the $ \rho-\mathrm{exchange} $ piece is not attractive enough in the domain of energy considered here to counteract this feature. For instance we know that this conclusion no longer holds in the longitudinal channel where the $ \pi-\mathrm{exchange} $ is attractive enough to overcome the Landau-Migdal interaction and to create a collective mode (the \textit{pionic branch} \cite{chanfray/ericson}). But, as we already see from eq. (\ref{eq:1}), the spin-isospin longitudinal channel is suppressed with regard to the transverse one and it plays a minor role in the neutrino-nucleus reactions.\\
The effect of the RPA correlations are less strong in the others partial channels and are not shown here. The resonant Delta channel (short dashed curve) arises at high energy transfer ($ \omega \sim 450 $ MeV).\\ 
The most intereting feature of the cross section is the importance of the (\textit{np-nh}) channel (dotted curve). We see that it gives a rather large contribution to the inclusive cross section and has an extended spectrum from the quasi-elastic to the Delta peaks. In particular these processes are important in the {\it dip} region, as mentionned previously.\\
The conclusion is that the inclusive neutrino-oxygen cross section is strongly modified with respect to the "bare" quasi-elastic one, which is often the sole channel entering into the simulations. In particular we have seen the occurence of large contributions from the two- and three-body channels. 

\section{Neutrino-Oxygen events yields}  
     
\begin{figure}[ht]
\begin{center}
\epsfig{figure=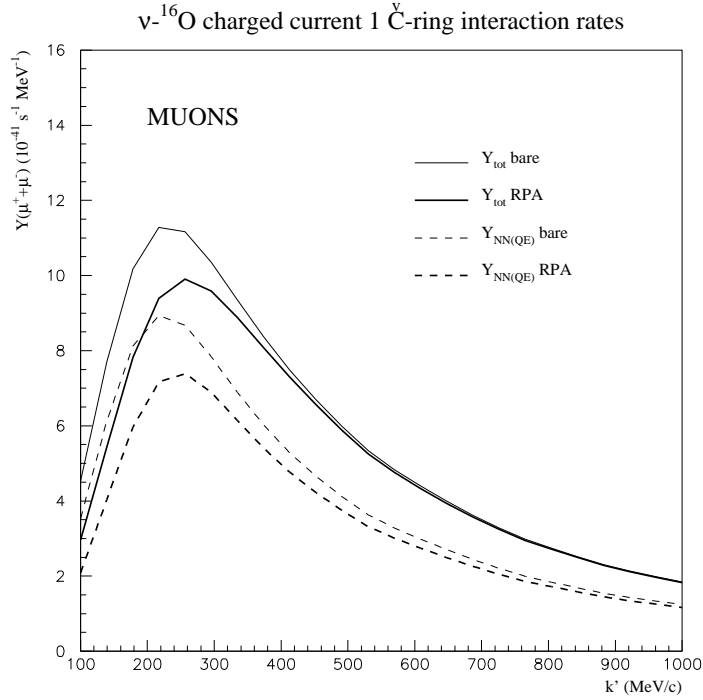,width=10cm,height=10cm}
\end{center}
\caption{One \v{C}erenkov $ \nu_\mu + ^{16}\hbox{O} $ events yields versus the muon momentum. The full curves correspond to the total 1 \v{C}.R. events yields with (thick curve) and without (thin curve) RPA. The dashed curves correspond to the quasi-elastic 1 \v{C}.R. events yields with (thick curve) and without (thin curve) RPA.
\label{fig:2}}
\end{figure}

The interesting quantities are the neutrino-oxygen events yields, which we compute at fixed charged lepton momentum:
\begin{equation} \label{eq:3}
Y(\nu_\alpha + \bar{\nu}_\alpha)(k^\prime) = \int_{E_{k^\prime}}^{\infty} \, dE \, \left( \Phi_{\nu_\alpha}(E) \, \frac{{\partial\sigma}(\nu- ^{16}\hbox{O})}{\partial k^\prime}(E,k^\prime) \, + \, \Phi_{\bar{\nu}_\alpha}(E) \, \frac{{\partial\sigma}(\bar{\nu}- ^{16}\hbox{O})}{\partial k^\prime}(E,k^\prime) \right) 
\end{equation}
where $ E $ is the neutrino energy, $ \Phi_\nu $ the incoming neutrinos flux and $ \partial\sigma/\partial k^\prime $ the neutrino-oxygen cross section computed in the previous section. We use the fluxes of Bartol \cite{barr/gaisser/stanev} in our calculations. The main feature of these fluxes is their sharp decrease with increasing neutrino energy. Within the few rough assumptions described previously, we classify our events with the number of \v{C}erenkov rings they produce. The results for the 1 \v{C}.R. events yields, which are relevant in the atmospheric neutrino experiments, are shown on fig. (\ref{fig:2}) for incident $ \nu_\mu $ and $ \bar{\nu}_\mu $, the full curves corresponding to the total 1 \v{C}.R. events yields and the dashed curves to the sole quasi-elastic 1 \v{C}.R. events yields. We give the results of the calculations without (thin curves) and with (thick curves) RPA.\footnote{Let us note that in order to give a full calculation of the events rates, one has to include the neutral currents. In this case, because the scattered neutrino is not seen experimentally, the resonant and coherent channels belong to the required 1 \v{C}.R. class (assuming that the pion is above threshold). The calculations have been performed but the analysis needs more informations on detection efficencies and on absorption of pions. This problem has to be investigated further.} 
 
First we observe that the RPA tends to reduce the events yields. Indeed it shifts the cross section towards high energies which are disfavored by the neutrino flux. The maximal reduction factor is of the order of 10 \%. As expected, the (\textit{np-nh}) excitations on the contrary increase the absolute events yields. At the maximum value of the yields, the enhancement of the total yield with respect to the "bare" quasi-elastic one is around 20 \%. This result reflects the main features of the cross sections. Furthermore one must be aware that this result is just a lower limit of the true enhancement. Indeed if a pion produced through Delta decay or coherent emission is absorbed in the medium, it will not produce a \v{C}erenkov ring and therefore the whole corresponding charged-current process belongs to the 1 \v{C}.R. events class. This problem remains to be investigated. Nevertheless we can conclude that the neglected reactions of the atmospheric neutrinos on pairs of correlated particles in an oxygen nucleus, which lead to 1 \v{C}.R. events, are a possible explanation for the under-estimation of the experimental $\nu_e + \bar{\nu_e}$ data by the Monte-Carlo simulations in Super-Kamiokande \footnote{We thank Pr. Y. Declais who caught our attention on this point.}. Furthermore we insist on the fact that these modifications to the theoretical events rates could have rather large consequences on the results of the experimental collaborations which are now based on the {\it absolute rates} of events and not only on the ratios of these events rates.\\

\section{Conclusion}     
     
In this work we studied the neutrino-oxygen reactions in the domain of energy relevant for the atmospheric neutrinos experiments using large underground \v{C}erenkov detectors. First we achieved the calculations of the nuclear responses of oxygen to a neutrino excitation. The semi-classical model used here has been developped to study the nuclear intermediate energy spectrum, from the quasi-elastic to the Delta peaks. It includes a parameterization of the multi-particles excitations ({\it np-nh}, {\it n = 2,3}) and the possible interactions between excited states by the means of the ring approximation to the RPA.\\
Within a few assumptions we perform an analysis of our results with respect to the number of \v{C}erenkov rings produced in the various nuclear processes considered. We conclude first that the RPA correlations tend to reduce the quasi-elastic and therefore the total {\it 1 \v{C}erenkov ring} events yields, the maximal reduction factor being of the order $\sim$ 10 \%. 
These events are those retained by the experimental cuts performed in the Kamiokande, IMB and Super-Kamiokande collaborations, whose results are our main point of interest. The second result is that the (\textit{np-nh}) excitations increase the total {\it 1~\v{C}erenkov~ring} events yields with respect to the "bare" quasi-elastic one usually taken into account. The strong enhancement factor ($\sim$ 20 \%) could lead to modifications in analyses of experimental results based on {\it absolute} number of events.

\section*{Acknowledgments}     
We wish to gratefully acknowledge G.Chanfray, Y.Declais, P.Lipari and S.Katsanevas for stimulating and enlightening discussions. 
 
\newpage

\section*{References}     

\newpage

\section*{Appendix}     

\begin{figure}[hb]
\begin{center}
\epsfig{figure=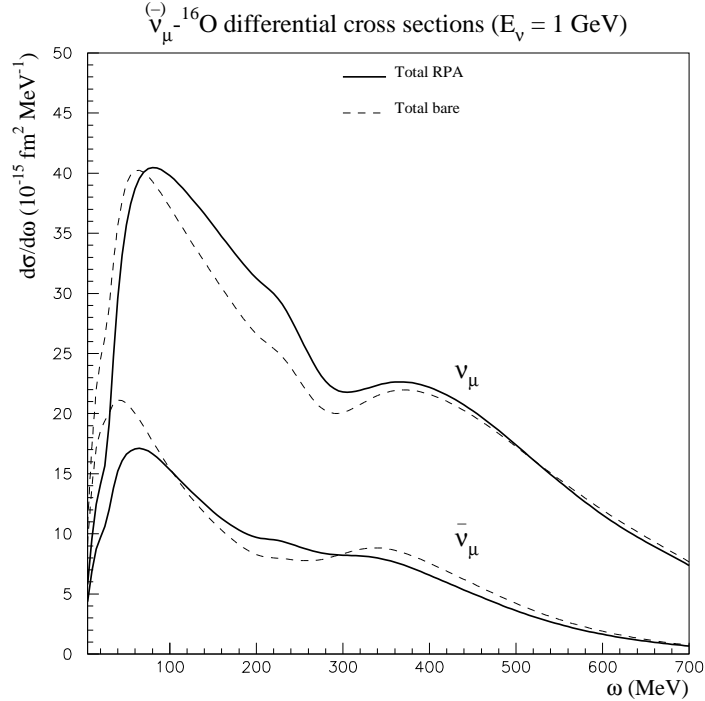,width=10cm,height=10cm}
\end{center}
\caption{$ \nu_\mu (\bar{\nu}_\mu) + ^{16}\hbox{O} $ differential cross sections
versus the energy transfer. The full thick (dashed thin) curves correspond to 
the RPA (bare) calculations.
\label{fig:3}}
\end{figure}

\begin{figure}[H]
\begin{center}
\epsfig{figure=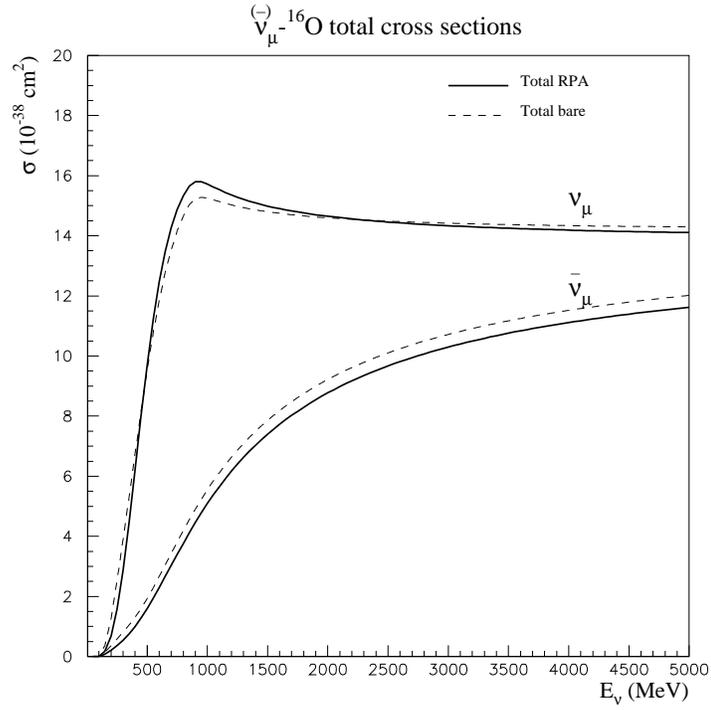,width=10cm,height=10cm}
\end{center}
\caption{$ \nu_\mu (\bar{\nu}_\mu) + ^{16}\hbox{O} $ total cross sections
versus the neutrino energy. The full thick (dashed thin) curves correspond to 
the RPA (bare) calculations.
\label{fig:4}}
\end{figure}

\begin{figure}[H]
\begin{center}
\epsfig{figure=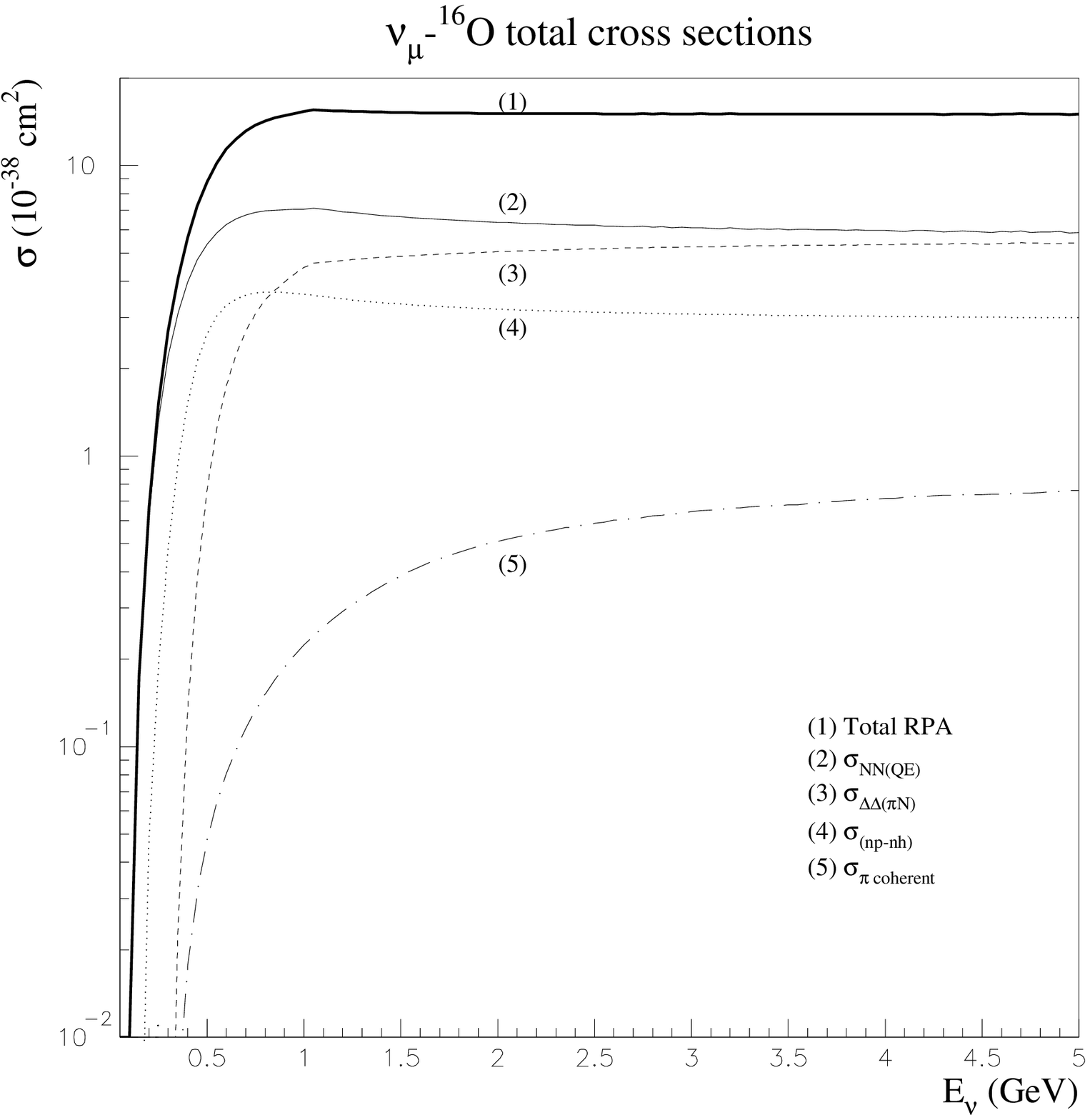,width=10cm,height=10cm}
\end{center}
\caption{$ \nu_\mu + ^{16}\hbox{O} $ total RPA cross sections
versus the neutrino energy. The full thick curve correspond to 
the inclusive reaction channel which is the sum of the following partial
channels: quasi-elastic (full thin curve), Delta resonant (dashed curve),
$np-nh$ (dotted curve), coherent pion emission (dot-dashed curve).
\label{fig:5}}
\end{figure}

Effects of the RPA on the differential (figure \ref{fig:3}) and total (figure \ref{fig:4}) 
$ \nu_\mu (\bar{\nu}_\mu) + ^{16}\hbox{O} $ cross sections. The total cross sections are obtained by numerical integration
of the differential ones $\partial\sigma / \partial k^\prime $ on the lepton momentum. The thick curves 
display the RPA cross sections while the thin dashed curves represent the 
bare cross sections. The $ \bar{\nu}_\mu + ^{16}\hbox{O} $ are the most 
sensitive reaction channels to the RPA effects. This is true for the whole range
of neutrino energy considered here and results in a quenching of the cross
sections which is less than 10 \%.\\
The last figure (figure \ref{fig:5}) shows the various contributions to the inclusive total cross sections. The RPA
responses ar used to evaluate the cross sections. The most important cross sections are the quasi-elastic and the resonant
Delta ones (respectively full thin and dashed curves). The rather large contribution of the $np-nh$ reaction channels 
(dotted curve) is obviously seen on the figure. Finally the coherent pion emission channel (dot-dashed curve) tends to
increase with the neutrino energy. However one has to notice that the contribution of this channel remains rather weak
(typically one order of magnitude below the previous contributions).   

\newpage

\begin{thebibliography}{99}

\bibitem{kamiokande}
Y.Fukuda \textit{et al.}, Phys. Lett. B\textbf{335}, 237(1994).

\bibitem{superkamiokande}
Y.Fukuda \textit{et al.}, Phys. Lett. B\textbf{433}, 9(1998); Phys. Rev. Lett. \textbf{81}, 1562(1998).

\bibitem{imb}
R.Becker-Szendy \textit{et al.}, Nucl. Phys. B\textbf{38}, 331(1995).

\bibitem{soudan}
W.W.M.Allison \textit{et al.}, Phys. Lett. B\textbf{391}, 491(1997).   

\bibitem{barr/gaisser/stanev}
G.Barr, T.K.Gaisser, T.Stanev, Phys. Rev. D\textbf{39}, 3532(1989).

\bibitem{gaisser/honda/etal}
T.K.Gaisser \textit{et al.}, Phys. Rev. D\textbf{54}, 5578(1996). 

\bibitem{marteau}
J.Marteau, to appear in European Physical Journal A; hep-ph 9902210, LYCEN 9905.

\bibitem{delorme/guichon}
P.A.M.Guichon, J.Delorme, \textit{Journ\'ees d'\'etudes de Saturne}, Piriac (1989);\\
J.Delorme, P.A.M.Guichon, Phys. Lett. B \textbf{263}, 157(1991).

\bibitem{oset/salcedo}
E.Oset, L.L.Salcedo, D.Strottman, Phys. Lett. \textbf{165}B, 13(1985). 

\bibitem{shimizu/faessler}
K.Shimizu, A.Faessler, Nucl. Ph. A\textbf{333}, 495(1980). 

\bibitem{chanfray/ericson}
G.Chanfray, M.Ericson, Phys. Lett. B\textbf{141}, 163(1984).

\end{thebibliography}
\end{document}